\documentclass[sigconf,10pt]{acmart}
\renewcommand\footnotetextcopyrightpermission[1]{} %
\usepackage{physics}
\usepackage{amsmath}
\usepackage[utf8]{inputenc}
\usepackage{natbib}
\usepackage{graphicx}
\usepackage{tikz}
\usetikzlibrary{quantikz}
\usetikzlibrary{arrows}
\usepackage{adjustbox}
\usepackage{balance}

\begin{document}

\title{Quantum Computing for Location Determination}

\author{Ahmed Shokry}
\affiliation{%
  \institution{Alexandria University}
  \city{Alexandria}
  \country{Egypt}}
\email{ahmed.shokry@alexu.edu.eg}

\author{Moustafa Youssef}
\affiliation{%
  \institution{AUC and Alexandria University}
  \city{Alexandria}
  \country{Egypt}}
\email{moustafa@alexu.edu.eg}

\begin{abstract}
Quantum computing provides a new way for approaching problem solving, enabling efficient solutions for problems that are hard on classical computers. It is based on leveraging how quantum particles behave. With researchers around the world showing quantum supremacy and the availability of cloud-based quantum computers with free accounts for researchers, quantum computing is becoming a reality.

In this paper, we explore both the opportunities and challenges that quantum computing has for location determination research. Specifically, we introduce an example for the expected gain of using quantum algorithms by providing an efficient quantum implementation of the well-known RF fingerprinting algorithm and run it on an instance of the IBM Quantum Experience computer. The proposed quantum  algorithm  has a complexity that is exponentially better than its classical algorithm version, both in space and running time. We further discuss both software and hardware research challenges and opportunities that researchers can build on to explore this exciting new domain. 
  
\end{abstract}
\settopmatter{printacmref=false}

\keywords{quantum computing, location determination systems, quantum  sensors, quantum location determination, next generation location tracking systems}

\maketitle
\pagestyle{plain}

\section{Introduction}
Quantum Computing (QC) is a new field at the intersection of physics, mathematics, and computer science. It leverages the phenomena of quantum mechanics to improve the efficiency of computation. Specifically, to store and manipulate information, quantum computers use quantum bits (qubits). Qubits are represented by subatomic particles properties like the spin of electrons or polarization of photons. Quantum computers leverage the quantum mechanical phenomena of superposition, entanglement, and interference to create states that scale exponentially with the number of qubits, potentially allowing solving problems that are traditionally hard to solve on classical computers \cite{qbible}.

In 1994, Peter Shor presented a theoretical quantum algorithm that could efficiently break the widely used RSA  encryption algorithm \cite{shor}. In 1996, Lov Grover developed a quantum algorithm that dramatically sped up the solution to unstructured data searches from $o(n)$ to $o(\sqrt{n})$ \cite{grover}. Since then, interest in QC has sparked with a number of big companies and startups investing in realizing quantum computers and providing tools for programmers to develop quantum algorithms. This materialized in currently having real cloud-based quantum computers with free accounts to researchers \cite{ibm64,ibm_topologies} as well as a large number of quantum programming languages and simulators \cite{quantum_race,ibm64,google_supremacy}. In October 2019, Google announced that it has reached quantum supremacy with an array of 54 qubits by performing a series of operations in 200 seconds that would take a supercomputer about 10,000 years to complete~\cite{google_supremacy}. In December 2020, a Chinese research group reached quantum supremacy by implementing Boson sampling on 76 photons with the Jiuzhang quantum computer \cite{chinese_supremacy}. The quantum computer generates the samples in 20 seconds that would take a classical supercomputer 600 million years of computation.

In this paper, we explore the opportunities and challenges of applying quantum computing to the field of location determination. Specifically, we discuss both algorithmic and hardware advantages that QC provides for use with location determination as well as highlight the research challenges that need to be addressed to fully leverage their potential. As an example, we present a quantum RF fingerprint matching algorithm that requires space and runs in $o(m \log(N))$ as compared to the classical algorithm that requires space and runs in $o(m N)$, where $N$ is the number of access points and $m$ is the number of fingerprinting locations. This exponential speedup in complexity and saving in space can be further enhanced to $o(log(m N))$ using more advanced algorithms. We validate our algorithm on an instance of the IBM Quantum Experience platform and discuss its performance. We end the paper by a discussion of the different opportunities offered and challenges posed by quantum  computing to the field of location tracking systems.

The rest of the paper is organized as follow: we start with a brief background on quantum computing in Section~\ref{sec:background}. Section~\ref{sec:qfp} provides the details of our quantum fingerprint matching algorithm and its evaluation. We present different challenges and opportunities of QC for location tracking in Section~\ref{sec:challenges}. Finally, sections \ref{sec:related} and \ref{sec:conclude} discuss related work and conclude the paper, respectively.

\section{Background}
\label{sec:background}
In this section, we give a brief background on the basic concepts of quantum computing that we will build on in the rest of the paper.

\begin{figure}[t!]
	\centering
\begin{quantikz}
	\ket{0}  \,  & \gate{H} & \meter{} & \cw  \rstick[style={xshift=-10cm}]{$ \left\{ \begin{array}{rl}
			0 & \text{with prob. 0.5} \\
			1 & \text{with prob. 0.5} 
		\end{array} \right.$}
\end{quantikz}	
	\caption{A simple quantum circuit. Single lines carry quantum information while double lines carry classical information.}
	\label{fig:simplecircuit}
\end{figure}
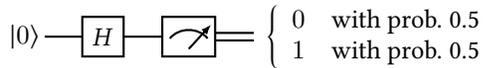

A quantum bit (qubit) is the basic unit of information and is analogue to the classical bit. Contrary to classical bits, a qubit can exist in a \textbf{superposition} of the zero and one states.  This superposition is what allows quantum computations to work on both states at the same time. This is often referred to as quantum parallelism. Qubits can have various physical implementations, e.g. the polarization of photons. 

Formally, the Dirac notation is commonly used to describe the state of a qubit as $\ket{\psi} = \alpha \ket{0} + \beta \ket{1}$, where $\alpha$ and $\beta$ are complex numbers called the amplitudes of classical states $\ket{0}$ and $\ket{1}$, respectively. The state of the qubit is normalized, i.e. $\alpha^2 + \beta^2 =1$. When
the state $\ket{\psi}$ is measured, only one of  $\ket{0}$ or $\ket{1}$ is observed, with probability  $\alpha^2$ and $\beta^2$, respectively. The measurement process is destructive, in the sense that the state collapses to the value $\ket{0}$ or $\ket{1}$ that has been observed, losing the original amplitudes  $\alpha$ and $\beta$ \cite{qbible}.

Operations on qubits are usually represented by gates, similar to a classical circuit.
 An example of a common quantum gate is the NOT gate (also called Pauli-X gate) that is analogous to the not gate in classical circuits. In particular, when we apply the NOT gate to the state $\ket{\psi_0} = \alpha \ket{0} + \beta \ket{1}$, we get the state $\ket{\psi_1} =  \beta \ket{0} + \alpha \ket{1}$. Gates are usually represented by unitary matrices 
while states are represented by column vectors\footnote{The ket notation $\ket{.}$ is used for column vectors while the bra notation $\bra{.}$ is used for row vectors.}. The matrix for the NOT gate is
$\begin{bmatrix}
	0 & 1\\
	1 & 0
\end{bmatrix}
$ and the above operation can be written as $\ket{\psi_1} = NOT(\ket{\psi_0}) = \begin{bmatrix}
	0 & 1\\
	1 & 0
\end{bmatrix} \begin{bmatrix}
\alpha\\
\beta
\end{bmatrix}.
$

Another important gate is the Walsh–Hadamard gate, $H$, that maps $\ket{0}$ to $\frac{1}{\sqrt{2}} (\ket{0}+ \ket{1})$, i.e. a superposition state with equal probability for $\ket{0}$ and $\ket{1}$; and maps $\ket{1}$ to $\frac{1}{\sqrt{2}} (\ket{0}- \ket{1})$. Figure~\ref{fig:simplecircuit} shows a simple quantum circuit. Single lines carry quantum information while double lines carry classical information (typically after measurement). The simple circuit applies an $H$ gate to state $\ket{0}$, which produces the state $\frac{1}{\sqrt{2}} (\ket{0}+ \ket{1})$ at the output of the gate. The measurement step produces either 0 or 1 with equal probability (the squared amplitude of the measured state). The state collapses to the observed classical bit value. 

It is important to note that the concept of quantum \textbf{interference} is at the core of quantum computing. Using quantum interference, one uses gates to cleverly and intentionally bias the content of the qubits towards the needed state, hence achieving a specific computation result. 

The notion of qubit can be extended to higher dimensions using a quantum register.
A quantum register $\ket{\psi}$, consisting of $n$ qubits, lives in a $2^n$-dimensional
complex Hilbert space $\mathcal{H}$. Register $\ket{\psi} = \sum_{0}^{2^n-1} \alpha_i \ket{i}$ is specified by complex numbers $\alpha_0, ..., \alpha_{2^n-1} $, where  $\sum |\alpha_i|^2 =1$. Basis state $\ket{i}$ denotes the binary encoding of integer $i$. We use the tensor product $\otimes$ to compose two quantum systems. For example,
we can compose the two quantum states  $\ket{\psi} = \alpha \ket{0} + \beta \ket{1}$ and  $\ket{\phi} = \gamma \ket{0} + \delta \ket{1}$ as $\ket{\omega} = \ket{\psi} \otimes \ket{\phi}= \alpha \gamma \ket{00} + \alpha \delta \ket{01}+ \beta \gamma \ket{10}+ \beta \delta \ket{11}$. %

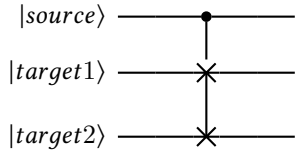
\begin{figure}[t!]
	\centering
	\begin{quantikz}
		\lstick{$\ket{source}$} &\qw & \ctrl{1} & \qw & \qw\\ 
		\lstick{$\ket{target 1}$}  & \qw  & \swap{1} & \qw & \qw \\
		\lstick{$\ket{target 2}$} &\qw    &           \targX{}                  & \qw & \qw\\	
	\end{quantikz}	
	\vspace{-0.7cm}
	\caption{An example of controlled gates. The two target qubits are swapped, if and only if, the source line is $\ket{1}$.} %
	\label{fig:controlled}
\end{figure}

Gates can also be defined on multiple qubits. For example,
Figure~\ref{fig:controlled} illustrates a frequently encountered gate in quantum circuits, the control gate. In a control gate, the operation (e.g. Swap) is performed on the target wire(s), if and only if, the source line is $\ket{1}$. This can be used to ``\textbf{entangle}'' qubits together. Entangled qubits are correlated with one another, in the sense that information on one qubit will reveal information about the other unknown qubit, even if they are separated by large distance \cite{qbible}.

A common way to describe a quantum algorithm is to use a quantum circuit, which is a combination of the quantum gates (e.g as in Figure~\ref{fig:simplecircuit}). The input to the circuit is a number of qubits (in quantum registers) and the gates act on them to change the combined circuit state using superposition, entanglement, and interference to reach a desired output state that is a function of the algorithm output. The final step is to measure the output state(s),  which reveals the required information.

Finally, the no-cloning theorem \cite{no_cloning} indicates that, counter to classical bits, quantum bits cannot be cloned. Therefore, one cannot assume that a quantum bit can be copied as needed (i.e. there is no fan-out as in classical circuits). This has a number of implications on designing quantum algorithm. For example, the no cloning theorem is a vital ingredient in quantum cryptography as it forbids eavesdroppers from creating copies of a transmitted quantum cryptographic key. On contrast, it makes classical error correction techniques not suitable for quantum states \cite{qbible}.

\section{A Quantum Fingerprint Matching Algorithm}
\label{sec:qfp}
In this section, we present a quantum version of the commonly used fingerprint matching algorithm. Fingerprinting-based location determination systems are used with RF and WiFi-based localization to capture the relation between the received signal strength (RSS) and user location \cite{youssef2005horus,bahl2000radar,cos_sim1, cos_sim2}. The idea is to collect the RSS signature/fingerprint of the APs in the area of interest at different discrete locations during an offline calibration phase. During the online location tracking phase, the received RSS from the different APs is compared to the fingerprint and the location in the fingerprint closest to the heard signal is returned as the estimated location. There are different similarity measures for matching the RSS vectors. One of the approaches that has been used to counter device heterogeneity is to use cosine similarity \cite{cos_sim1, cos_sim2}. 

We start the section by presenting a general quantum fingerprinting matching algorithm based on cosine similarity followed by a detailed example on how it works in a simple setting. We then quantify the performance of the algorithm by implementing it on a real quantum computer and discuss its efficiency and different aspects of its operation.

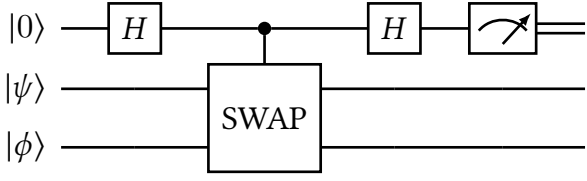
\begin{figure}[t!]
	\centering
	\begin{adjustbox}{width=0.45\textwidth}
		\begin{quantikz}[row sep=0.1cm]
			\lstick{$\ket{0}$} & \gate{H} &\ctrl{1} & \gate{H} & \meter{} & \cw \\
			\lstick{$\ket{\psi}$} & \qw & \gate[wires=2]{\text{SWAP}} & \qw & \qw & \qw \\
			\lstick{$\ket{\phi}$} & \qw & & \qw & \qw & \qw\\	
		\end{quantikz}	
	\end{adjustbox}
	\vspace{-0.5cm}
	\caption{Quantum fingerprint matching circuit between the online RSS vector (mapped into qubit register $\ket{\psi}$) and a single fingerprint RSS vector (mapped into qubit register $\ket{\phi}$).}
	\label{fig:full_swap}
\end{figure}

\begin{figure*}[t!]
	\centering
	\begin{adjustbox}{width=0.9\textwidth}
		\begin{tikzpicture}[row sep=0.1cm]
			\node at (-3.2,0.3) {State Preparation};
			\node at (0.0,0.3) {Fingerprint Matching};
			\node at (3.5,0.3) {Measurement};
			\draw[red,line width=1.6pt,dashed] (-1.5,0) -- (-1.5,-4);
			\draw[red,line width=1.6pt,dashed] (1.4,0) -- (1.4,-4);
			\node at (-6.3,-0.6) {Ancilla};
			\node[text width=2cm] at (-6.5,-1.9) {Testing RSS vector};
			\node[text width=1.5cm] at (-6.1,-3.3) {FP RSS vector};
			\node at (0,-0.1) [anchor=north]{
				\begin{quantikz}
					\lstick{$\ket{0}$} &\qw & \gate{H} &\ctrl{1} & \gate{H} & \meter{} & \cw \rstick{Sim. score = $f(p_1)$}\\ 
					\lstick{$\ket{0}$} & \gate{U(2\times\arctan(\frac{b}{a}))}   & \qw  & \swap{1} & \qw & \qw & \qw \\
					\lstick{$\ket{0}$} & \gate{U(2\times\arctan(\frac{d}{c}))}  &\qw    &           \targX{}                  & \qw & \qw & \qw\\	
				\end{quantikz}	
			};	
		\end{tikzpicture}
	\end{adjustbox}
	\vspace{-0.5cm}
	\caption{A detailed example of the quantum fingerprint matching circuit using two APs only. The circuit shows the state preparation step, i.e. how to map the training RSS vector $(a, b)$ and testing classical RSS vector $(c, d)$ to the quantum states $a \ket{0} + b \ket{1}$ and $c \ket{0} + d \ket{1}$, respectively, starting from $\ket{0}$.} %
	\label{fig:swap2}
\end{figure*}
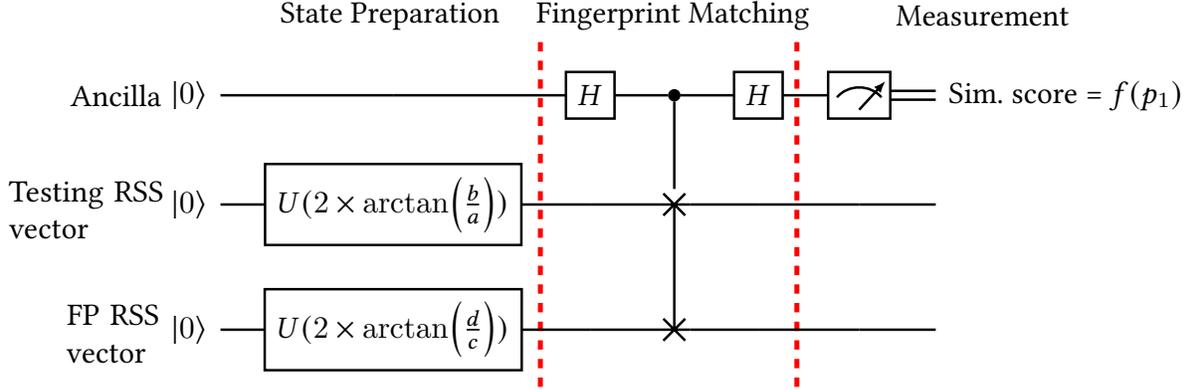

\subsection{General Quantum Cosine Fingerprint Matching Algorithm}
\label{sec:gen_alg}
Figure~\ref{fig:full_swap} shows the quantum circuit for calculating the cosine similarity between two normalized RSS vectors encoded in the quantum registers $\ket{\psi}$ (e.g. a test RSS vector during the online location tracking phase) and $\ket{\phi}$ (e.g. a single fingerprint RSS vector). In particular, the circuit calculates:
\begin{equation}
	\textrm{sim}(\ket{\psi}, \ket{\phi}) = \cos[2](\psi,\phi) = |\bra{\phi}\ket{\psi}|^2
\end{equation}

where $\cos(\psi,\phi)$ is cosine the angle between the  two normalized vectors $\psi$ and $\phi$ and $\bra{\phi}\ket{\psi}$ is the dot product between them.

Without loss of generality, assume an area of interest with $N= 2^n$ APs. Therefore, the N-dimensional normalized RSS vector from the $N$ APs $v= (\alpha_0, \alpha_1, ..., \alpha_{N-1}), \sum_{i=0}^{N-1}\alpha_i^2 =1$, can be encoded using an $n$ qubits register (notice the exponential saving in state) $\ket{\psi}$ (or $\ket{\phi}$) as $\ket{\psi} = \sum_{i=0}^{N-1} \alpha_i \ket{i}$, 
where the basis state $\ket{i}$ represents the binary encoding of integer $i$ \cite{qmemory,state_prep}. The general state preparation step for $n$ qubits, i.e. mapping a classical vector to a quantum register can be achieved efficiently using different quantum  circuits, e.g. \cite{qmemory,state_prep}. We give an example of how to prepare this state from classical vectors in the next section.

The input to the circuit consists of a single ancilla qubit and the two quantum registers encoding the RSS vectors, i.e. the input state is $\ket{0}\ket{\psi}\ket{\phi}$. First, we apply a Walsh-Hadamard gate ($H$) to the ancilla qubit ($\ket{0}$) to obtain the superposition state:
\begin{equation}	
\frac{1}{\sqrt 2} (\ket{0}\ket{\psi}\ket{\phi}+ \ket{1}\ket{\psi}\ket{\phi})
\end{equation}

Then, we apply the controlled-swap gate, which exchanges the two registers $\ket{\psi}$ and $\ket{\phi}$ if the state of the top wire is $\ket{1}$. %
Therefore, the combined system state becomes
\begin{equation}
\frac{1}{\sqrt 2} (\ket{0}\ket{\psi}\ket{\phi}+ \ket{1}\ket{\phi}\ket{\psi})
\end{equation}

 (note the swapping of the two registers in the second term). Applying the second $H$ gate on the top wire evolves the state to
\begin{equation}
\frac{1}{2} \ket{0}(\ket{\psi}\ket{\phi}+ \ket{\phi}\ket{\psi}) +  \frac{1}{2} \ket{1}(\ket{\psi}\ket{\phi}- \ket{\phi}\ket{\psi})
\end{equation} 

This state is the same as
\begin{multline}
\frac{1}{2} \left( \sqrt{2+2 |\bra{\psi}\ket{\phi}|^2} \ket{0}\frac{\ket{\psi}\ket{\phi}+ \ket{\phi}\ket{\psi}}{\sqrt{2+2 |\bra{\psi}\ket{\phi}|^2}}\right.  \\ 
+ \left. \sqrt{2-2|\bra{\psi}\ket{\phi}|^2}\ket{1}\frac{\ket{\psi}\ket{\phi}- \ket{\phi}\ket{\psi}}{\sqrt{2-2 |\bra{\psi}\ket{\phi}|^2}} \right)
\end{multline}
(by normalizing the joint state of the two quantum registers)
Finally, the probability of measuring the top (ancilla) qubit to be 1 is $\frac{1}{2} ( 1- |\bra{\phi}\ket{\psi}|^2)$, which is a function of the required similarity measure between the two vectors.

Practically, we repeat this circuit $k$ times to estimate the cosine similarity as $1- 2 \times \#\ket{1}/k$.

Note that the ancilla qubit is put into a superposition state then entangled with the other two quantum registers using the controlled swap gate.  The final Hadamard gate is selected so that it reflects the desired computation. %
 
\subsection{Example}
In this section, we illustrate the quantum fingerprint matching algorithm described in the previous section using a simple example of two vectors with two APs each. Let the two normalized RSS vectors to be matched be $v_1=(0.39, 0.92)$ and $v_2=(0.24, 0.97)$. The complete circuit for calculating the similarity between the two vectors is given in Figure~\ref{fig:swap2}. 

The circuit starts by preparing the states, i.e. mapping the RSS vectors $v_1$ and $v_2$ to the quantum equivalent $\ket{\psi}=0.39 \ket{0}+ 0.92 \ket{1}$ and $\ket{\phi}=0.24 \ket{0}+ 0.97 \ket{1}$, respectively. This is achieved by using the $U$ gate, where

 $U(\theta) = \begin{bmatrix}
	\cos(\theta/2) & -\sin(\theta/2) \\
	\sin(\theta/2) & \cos(\theta/2) 
\end{bmatrix}
$. The fingerprint matching part is the same as the one described in the previous section but operating on a single qubit for each vector (since we have only two APs). For the given example, the probability of measuring the first qubit to be in state 1 is %
0.014 ($\frac{1}{2} ( 1- |\bra{\phi}\ket{\psi}|^2)$) and hence the similarity score ($|\bra{\phi}\ket{\psi}|^2$) is 0.972. Note that this circuit has to be repeated for each of the $m$ fingerprint locations to determine its similarity score to the test RSS vector. The fingerprint location with the highest score is selected as the estimated user location.

\subsection{Implementation and Evaluation}
\begin{figure}[t!]
	\centerline
	{\includegraphics[width=0.45\textwidth]{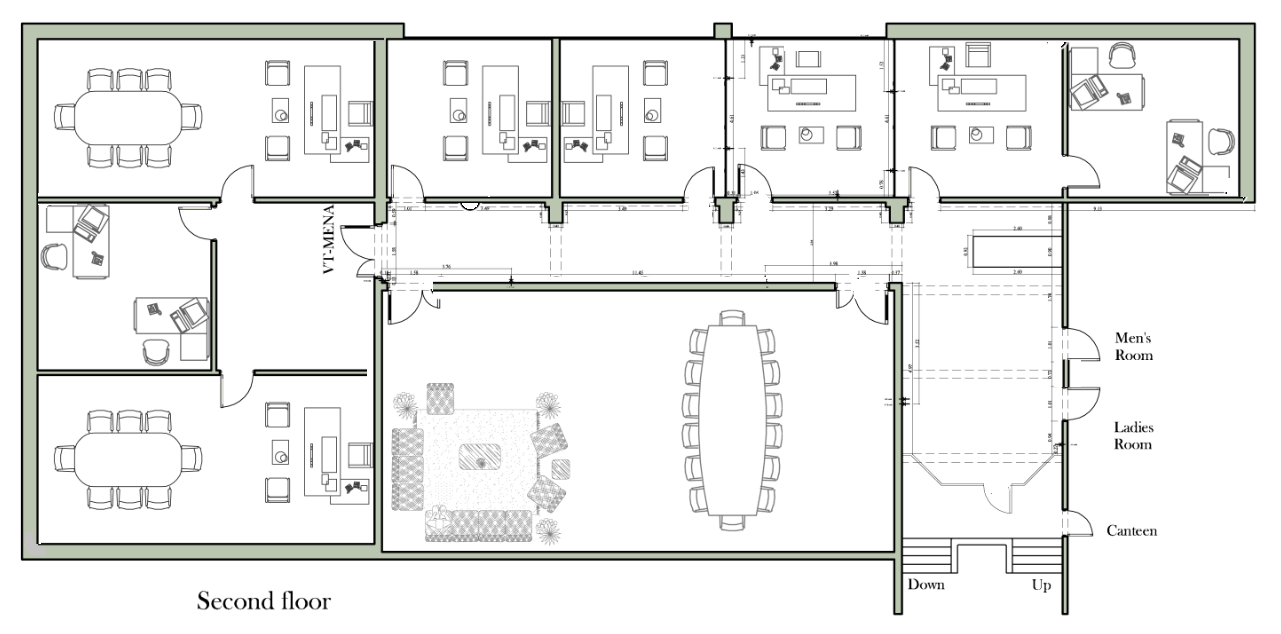}}
		\vspace{-0.5cm}
	\caption{Testbed used for experimental validation.}
	\label{fig:testbed}
\end{figure}

We deployed our system in a floor in our university campus building with a 89ft $\times$ 56ft %
area containing labs, offices, meeting rooms as well as corridors (Figure~\ref{fig:testbed}). We use the four already installed WiFi APs in the same floor as the testbed. We collect the WiFi scans by a Samsung S4 cell phone that scans for the WiFi access points at 24 different fingerprint locations. We also collected an independent test set at 24 other locations. Both the fingerprint and test locations are uniformly distributed over the entire area.

We implemented the full circuit in Figure~\ref{fig:full_swap} over the IBM Quantum Experience cloud platform \textit{ibmq\_athens} which has five qubits. This fits the number of APs in our testbed: two qubits for the four APs for the training  and testing locations (total of 4 qubits) and one ancilla qubit.

\begin{figure}[t!]
	\centerline
	{\includegraphics[width=0.45\textwidth]{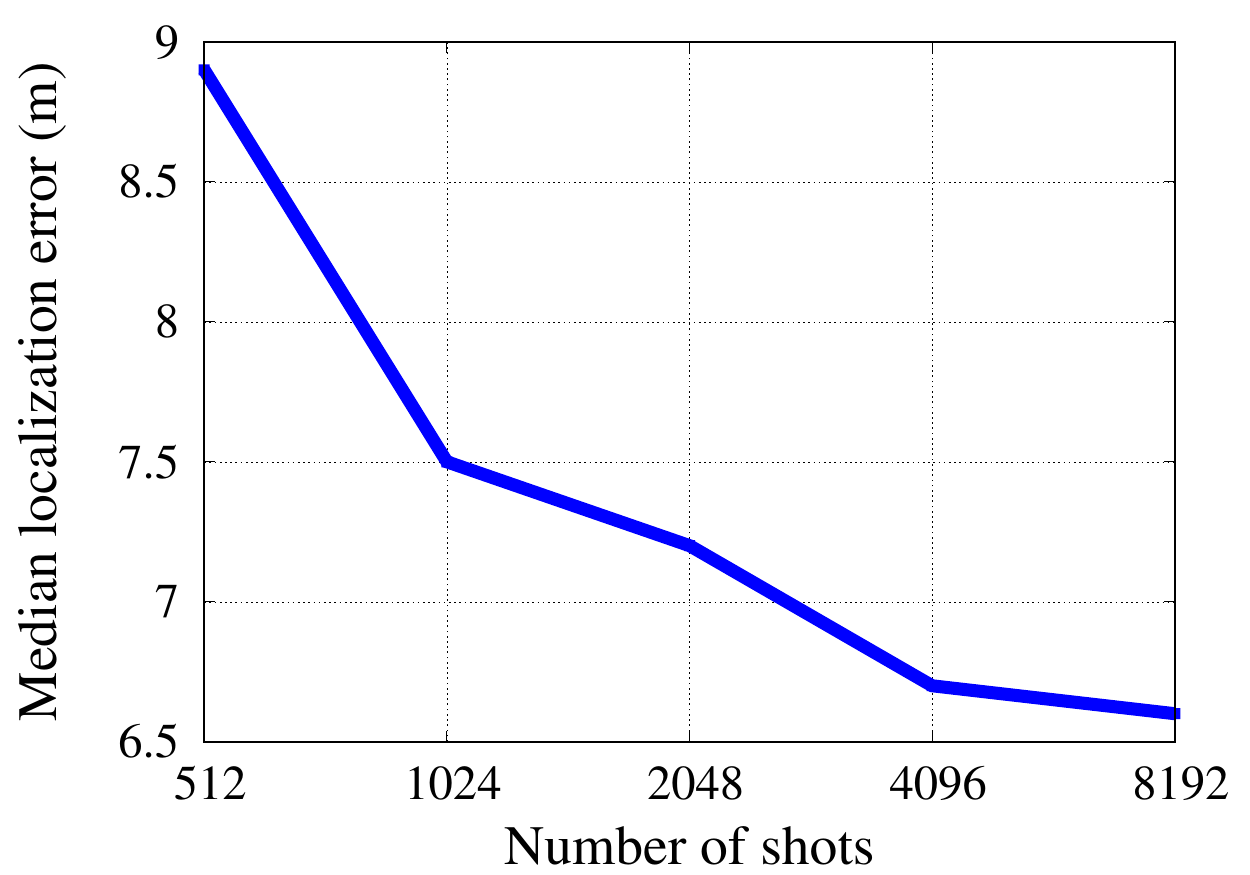}}
	\caption{Effect of changing number of shots (iterations) on the median localization accuracy.}
	\label{fig:shots}
\end{figure}

Figure~\ref{fig:shots} shows the effect of increasing the number of shots, i.e. re-running the system (parameter $k$ in Section~\ref{sec:gen_alg}), on the quantum system accuracy. The figure shows that, as expected, increasing the number of shots increases the system accuracy till it saturates around 4096 shots.

\begin{figure}[t!]
	\centerline
	{\includegraphics[width=0.45\textwidth]{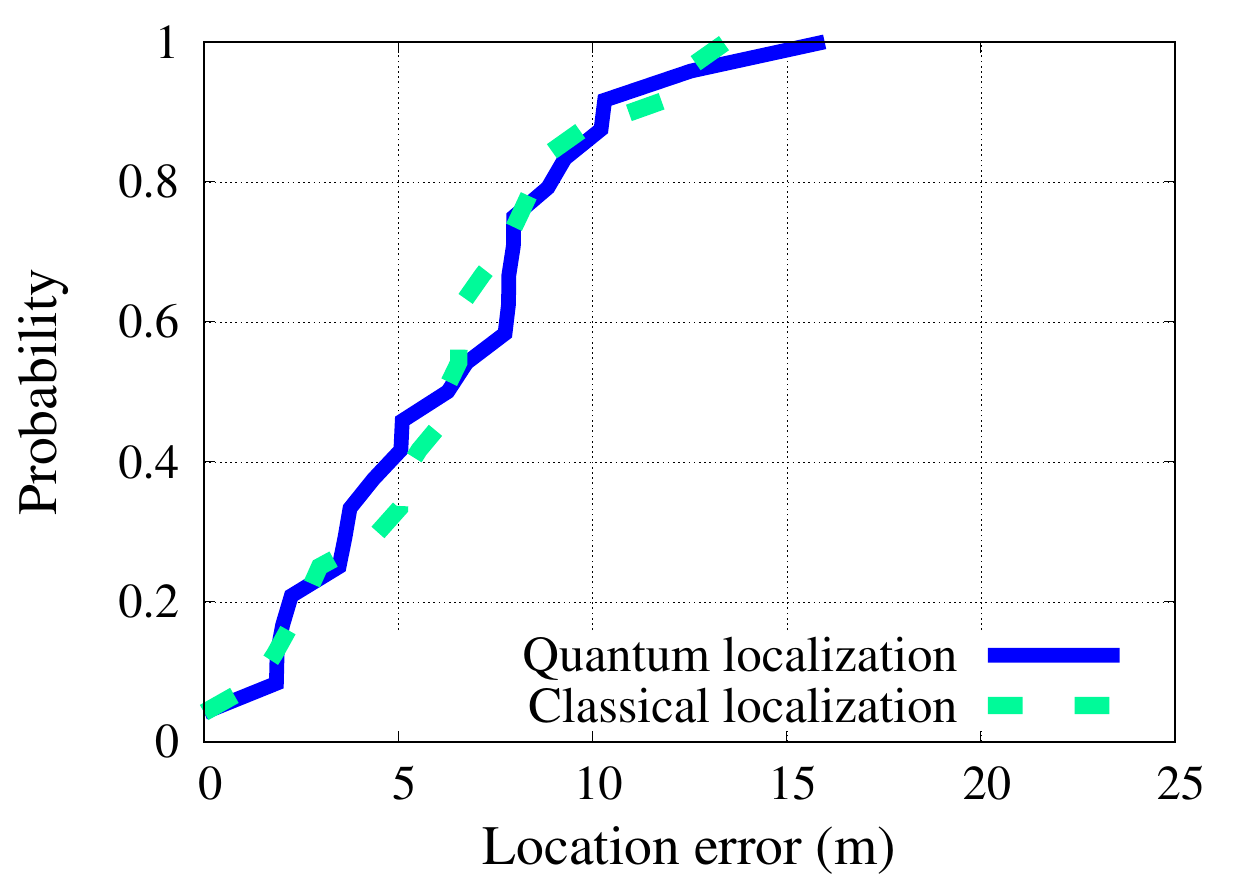}}
	\caption{CDF of localization error.}
	\label{fig:cdf}
\end{figure}

Figure \ref{fig:cdf} shows the CDF of distance error for the classical and the quantum fingerprinting localization system. The figure confirms that the two systems have the same performance, with the potential gain of quantum systems as we discuss in the next section.

\subsection{Discussion}
The classical version of this algorithm requires $o(N m)$ space and matching runs in $o(N m)$, where $m$ is the number of fingerprint locations and $N = 2^n$ is the number of APs. In contrast, the presented quantum algorithm in this section requires $o(m \log N)$ space and matching runs in $o(m \log N)$. This is an exponential enhancement in both space and run time.
 
The performance of the proposed algorithm can be further enhanced by encoding all the fingerprint locations using quantum states. This can lead to $o(\log (N m))$ running time. Furthermore, this can be combined with an idea similar to the Amplitude Amplification component of the Grover algorithm \cite{grover} to find the top matching fingerprint locations without re-running the circuit. This can significantly reduce the running time of the algorithm.

\section{Challenges and Opportunities}
\label{sec:challenges}
Building on the feasibility experiment we showed in the previous section, we now discuss some of the challenges and opportunities related to using quantum hardware and algorithms for location tracking.

\subsection{Quantum Sensors and New Ubiquitous Navigation Technologies}

Although GPS is considered a ubiquitous outdoor localization technology, it has a number of shortcomings: its accuracy severely degrades/becomes unavailable in urban canyons; it does not work in important environments such as indoors, deep under water, and in space; and it is susceptible to jamming attacks~\cite{gps_jamming}. It is estimated that five days of denial of the satellite service would cost the UK alone 5.2 billion pounds \cite{gps_denial_cost_uk}. There have been a number of GPS-replacement technologies over the years for mobile devices that mainly depend on fusing the different phone sensors, especially the inertial sensors \cite{gps_replace1,gps_replace2,dejavu}. However, traditional inertial sensors are noisy and the error in location estimation based on them accumulates quickly.

Recently, there have been an active research in developing quantum inertial sensors, e.g. quantum accelerometers \cite{quantum_acc_msquare}. For example, the work in \cite{quantum_acc_msquare,cool_atoms} proposes a high -precision and -accuracy quantum accelerometer that relies on  measuring properties of supercool atoms. A laser beam is used to measure the minute changes in the quantum wave properties of falling atoms as they respond to the object acceleration. The current system targets navigation of large vehicles. Nonetheless, it provides the feasibility of such system and opens the door for further miniaturization. %

These quantum accelerometers offer new opportunities for localization including being self-contained (need no access to external infrastructure), being not vulnerable to jamming attacks, providing both indoors and outdoors localization, and enabling accurate navigation in GPS-denied areas such as urban areas, space, and underwater.

Similarly, quantum magnetometers have been proposed \cite{quantum_magnetometer} that can measure both the earth's magnetic field strength and its direction accurately using synthetic diamonds special impurities and laser. The idea is that, when a laser beam is applied to the diamond, it emits light depending on the magnetic field that it is in. 
Systems that leverage these high-accuracy high-precision sensors can be developed based on dead-reckoning to estimate an object position autonomously. This can be further combined with traditional landmark-based localization techniques \cite{unloc} to further reduce error and provide anywhere high-accuracy location tracking.

The authors in \cite{entanglement_sensing} leverage entanglement of multiple sensors to achieve unprecedented sensitivity level in distributed RF sensing problems such as measuring the angle of arrival of an RF field. This in turn can be used to enhance angle-of-arrival based localization systems.

For 3D applications such as E911 and 3D navigation, researchers have developed a quantum pressure sensor that uses fundamental properties of helium atoms \cite{quantum_pressure}. The idea is that a laser beam going through a glass chamber filled with helium will change color depending on the air pressure. This can be used to measure the atmospheric pressure with high accuracy.

In 2020, the USA Defense Innovation Unit opened a call for a compact, high-performance sensor that can use quantum technology to provide precise inertial measurements in deep space\footnote{\url{https://www.diu.mil/}}. The call requires a prototype of such system before 2022 with error rates better than 100 meters per hour in deep space or 30 meters per hour for terrestrial applications while being no bigger than 0.1 cubic meters. This further highlights that quantum sensors will be available in the near future with the potential of being a game changer in localization. Nonetheless, these sensors provide both new opportunities and challenges for localization systems, both at the hardware and algorithmic levels.

\subsection{Quantum Device-free Localization} %
Device-free localization \cite{dfp} has been an active area of research. The idea is to analyze the change of the ambient RF signals to detect, track, and identify activities of events and objects in an environment, without them carrying any devices. Applications include intrusion detection, smart homes, gesture recognition, emotion detection, among many others.

With the introduction of different quantum sensors that allow higher accuracy, precision,  and sensitivity, new possibilities for device-free detection and sensing are unlocked. This includes, not only higher accuracy and range for the current applications, but also  new device-free applications and services. For example, one may use the entangled sensors described in the previous section for, e.g., human identity detection; a task that has proved to be difficult so far with the classical sensors. Similarly, a high-sensitivity quantum magnetometer may be used, in a device-free manner, to look inside vehicles to see if a suspicious object is inside, e.g. a bomb, or estimate the car speed and model. These can also be extended to the recent energy-free sensing concept \cite{solargest}.

\subsection{Rethinking Location Tracking Algorithms}
Given the potential of quantum computing algorithms and quantum sensors, this opens the door to rethink different location determination technologies and algorithms.

For example, given the exponential gain in performance of the quantum-based fingerprinting algorithm provided in this paper, one may wonder whether we still need to do clustering of fingerprint locations to reduce the computational requirements or not.

More drastically, given the expected high resolution and robustness of quantum inertial sensors, do we still need to have RF-based localization or a simple dead-reckoning localization system would be enough to provide both indoor and outdoor localization? Some may argue that even with a high precision sensors, dead-reckoning errors will still accumulate, though over longer periods. Therefore, some error resetting mechanism may still be needed, e.g. based on virtual landmarks in the environment \cite{unloc}. 

Similarity, one area that is open for contribution is how to implement different classical location determination algorithms on quantum computers. For example, it is not straightforward to generalize the cosine similarity measure used in this paper to other similarity measures, e.g. a probabilistic measure. 

Another area that has the potential to benefit from quantum computing is automatic construction of RF fingerprints. Typically, this has been performed in literature using ray tracing \cite{airadne,aroma}. The accuracy of these approaches are usually limited by the computational resources. The parallelism inherit in quantum computing may help provide higher-accuracy and faster ray tracing algorithm. Nevertheless, developing quantum algorithms for ray tracing is still an open area.

Given the potential of compressing the classical state by encoding them in quantum computers and the exponential gain in running time, quantum computing may provide an advantage to machine learning algorithms, especially those heavy on data such as deep learning systems. There are active research in the area of quantum machine learning \cite{QML1,QML2} and tools such as the Quantum Development Kit from Microsoft\footnote{\url{https://azure.microsoft.com/en-us/resources/development-kit/quantum-computing/}} and Google TensorFlow Quantum \cite{TFquantum} come with a
quantum machine learning library. Nonetheless, it is not clear if they fit the different location tracking sub-tasks and they may not be general to adapt to different measures and practical situations. %

\subsection{Secure Localization}
By secure localization we refer to problems including location privacy and location verification. For example, by using quantum senors, one can implement a stand-alone localization system that can run completely on the user device. This provides higher location privacy for the users.

Location verification is a well-studied classical problem \cite{dist_bounding1,dist_bounding2,dist_bounding3,dist_bounding4}, where a device wants to prove to a verifier that it is located in a specific area/location. This is typically performed by the distance bounding protocol, where the distance between the device and verifier is estimated based on time of flight, which is bounded by the speed of light, and a prover can only increase its distance to the verifier. Two of the common attacks on the distance bounding algorithm is  reply attack and the wormhole attack, both based on the attacker intercepting the messages and replaying them. 
Given that the quantum state cannot be copied, according to the no cloning theorem \cite{qbible}, quantum algorithms may provide a more secure version of location verification that can  counter classical attacks. Designing and analyzing location veri fication and attack schemes using quantum algorithm is an open research area.

\subsection{Theoretical Analysis}
In this paper, we showed how quantum algorithms can have an advantage over classical algorithms in terms of required storage and running time. An active area of research in quantum computing is the quantum complexity theory that studies the complexity classes of quantum algorithms and their relation to  classical counterparts \cite{qtheory}. For example, the class of bounded error, quantum, polynomial time (BQP) studies the problems that can be solved efficiently by a quantum computer with bounded error. Analyzing the complexity of the developed quantum algorithms for location tracking systems and obtaining performance bounds under different quantum complexity models is an important research direction.

 \subsection{Quantum Computing Resources}
With the promise and advantages offered by quantum computing algorithms, there has been a large interest of big players and startups to invest in developing quantum computing resources and making them available for researchers. In particular, there is a number of actual cloud quantum computers, quantum programming languages, and quantum simulators that enable researchers to experiment with QC and develop their algorithms.

We have used the IBM Quantum Experience Cloud Computers to implement the quantum fingerprint matching algorithm in this paper. Currently, IBM has a 64 volume\footnote{The quantum volume is a metric that combines the number of qubits, error rate, and topology in a single number.} quantum computer based on superconducting transmon qubits with a plan to double its quantum volume each year \cite{ibm64}.  Intel Labs is using  silicon spin qubits to achieve production-level quantum computing within ten years \cite{intel_qc}. Google claimed quantum supremacy in 2019 using its 54 qubits Sycamore processor~\cite{google_supremacy}. Honeywell, using  ion traps quantum  processors, claimed to produce the current most powerful quantum processor with a quantum volume of 128 \cite{honeywell}. Other startup companies; e.g. IonQ, Rigetti, and D-Wave; provide different approaches to realizing quantum computers including superconducting qubits, ion traps, and quantum annealing, respectively.  Microsoft has plans to develop its own topological qubit-based quantum computers \cite{quantum_race}. It also teamed up with other companies, including Honeywell and IonQ, to provide Quantum Azure, a quantum version of its cloud computing platform. Amazon is providing its Amazon Braket cloud platform, as part of its Amazon Web Services, that will run on top of different quantum computing hardware including IonQ, Rigetti, and D-Wave.

These platforms also provide tools to create and simulate quantum software, including IBM's Qiskit, Microsoft's Quantum Development Kit and Google's Cirq tool. Programming them is based on libraries in existing languages, such as Python, or new languages, such as Q\# from Microsoft.

All these initiatives highlight the promise of quantum computing and call for developing quantum location tracking algorithms libraries that can speed up the adoption of quantum computing algorithms for localization.

\subsection{Quantum Software/Hardware Interaction}
Due to the nascent nature of the current quantum computers, they are limited in resources and functionality, e.g. the number of qubits they can handle, the interconnection between qubits, what gates can be performed, and what gates can be performed on what qubit in a certain quantum computer. In addition, there are constraints on the error rates and decoherence time, i.e. loss of the quantum properties, and the gate timing. For example, the IBM  Quantum Experience Computers have different topologies for qubits interconnections \cite{ibm_topologies}. Therefore, not all gates can be applied to all qubits. For instance, the controlled NOT gate requires operation on two qubits. Therefore, these two qubits need to be directly connected in the topology. Otherwise, an extra swap operation needs to be added to allow this. %

The algorithm designer needs to be aware of that when developing her algorithm, which adds another layer of complexity to design efficient algorithms. Alternatively, quantum compilers may be developed that take these constraints into account and automatically convert an algorithm to fit a particular architecture.

\subsection{Energy Efficiency}

Energy-consumption is an important factor for the battery-operated mobile devices. Classical computers usually consume large amount of energy for computing and storage, generating considerable heat in the process. Quantum computers may offer a more energy-efficient way for computation. For example, the authors in \cite{magnet_comp} propose a circuit that can perform computations based on spin wave, which is a quantum property of electrons, in magnetic materials. Specifically, the circuit uses nanofilms of magnetic material to modulate the spin waves, without any electrical current. This can be leveraged in quantum computers to perform computation with virtually zero energy.

\subsection{QC for Mobile Computing and Communication}
Although we focused in this article on quantum computing for location tracking systems, the same concepts are also applicable to the fields of mobile computing and communication. Different algorithms can be revisited in the view of quantum computing for better efficiency. 

The quantum Internet, where quantum computing concepts such as entanglement are used to realize secure communication not achievable with traditional networks, are currently being developed with the availability of quantum satellites \cite{q_comm1,q_comm2}. Similarly, researchers have developed a system  based on quantum measurement of photons that can detect exceptionally weak signals  \cite{nist_faint_Signals}. This has the potential to decrease the energy requirement of traditional communication systems by orders of magnitude, or alternatively increase the communication range for the same energy budget. Another research effort has developed a single quantum sensor, based on using Rydberg atoms as electric field sensors, that can detect communication signals over the entire radio frequency spectrum from 0 to 100 GHz \cite{wide_range_spectrum}. 

Such research effort has the potential of revolutionizing the wireless communication field and its application as we know it.

\section{Related Work}
\label{sec:related}
In this section, we discuss both classical fingerprinting-based techniques as well as the general quantum-based algorithms.

\subsection{Classical Fingerprinting-based Localization Systems}
Classical fingerprinting-based RF localization systems use different matching functions to compare the online RSS vector to those stored in the fingerprint. These matching functions include Euclidean, Manhattan, Chi-Squared, Bray-Curtis, Mahalanobis, and cosine similarity \cite{bahl2000radar,cos_sim1, cos_sim2,del2009efficient, beder2012fingerprinting}. The last one is usually used to combat the device heterogeneity effects \cite{cos_sim1, cos_sim2}. Similarly, probabilistic techniques, based on the MLE estimator, e.g. \cite{youssef2005horus}; have also been introduced.

One key aspect common to all classical techniques is the need to compare the online measurements with the offline ones at each fingerprint location, making their time and space complexity $O(mN)$, where $m$ and $N$ are the number of fingerprint locations and APs, respectively.
\textit{On the other hand, quantum-based algorithms can provide an exponential advantage in both time and space as we presented in this paper.} Developing quantum algorithms that can provide the different similarity functions in literature is still an open area of research.

\subsection{Quantum Algorithms}
Quantum algorithms have a promise for significant performance gains in different fields, including scientific computing \cite{moller2017impact}, robotics \cite{petschnigg2019quantum}, cryptography \cite{mavroeidis2018impact, cheng2017securing}, chemistry \cite{primas2013chemistry, improta2016quantum}, finance \cite{rebentrost2018quantum}, among others. To allow this, researchers have developed quantum algorithms to tackle general mathematical problems like solving linear systems of equations \cite{subacsi2019quantum}, linear differential equations \cite{xin2020quantum}, searching for an element in unsorted list \cite{grover}, and integer factorization \cite{shor}. 

Quantum algorithms substantially outperform their classical counterparts. For example, \cite{harrow2009quantum}, provides a quantum algorithm for solving linear systems of equations, which is exponentially faster than classical algorithms tackling the same problem. The well-known Grover algorithm searches for entries in an unsorted database of size $n$ in $O(\sqrt{n})$ steps \cite{grover}. The Shor algorithm provides a polynomial time quantum algorithm for integer factorization based on a quantum Fourier transform sub-module \cite{shor}. \textit{In this paper, we provided a quantum fingerprinting matching that significantly outperforms its classical counterparts in both space and time. We also discussed different opportunities and challenges to further explore quantum algorithms in location tracking.}

\section{Conclusion}
\label{sec:conclude}
We presented a cosine-similarity quantum circuit for fingerprint matching in RF-based  localization systems. We implemented the proposed circuit on a physical IBM Quantum Experience cloud computer  and evaluated it in a real testbed. The proposed quantum algorithm provides an exponential enhancement of both the space and running time complexity. We further discussed different software and hardware  opportunities offered and challenges raised by using quantum computers for location determination. We hope this paper will excite location tracking researchers in particular and mobile wireless networking researchers in general to explore this emerging field.

\balance
\bibliographystyle{ACM-Reference-Format}
\bibliography{quantum}

\end{document}